# Design and Initial Performance of SHARP, a Polarimeter for the SHARC-II Camera at the Caltech Submillimeter Observatory

H. Li [a*], C. D. Dowell [b,c], L. Kirby [d], G. Novak [a], and J. E. Vaillancourt [c,e]

[a] Northwestern University, Dept. of Physics and Astronomy, Evanston, IL, 60208, USA
[b] Jet Propulsion Laboratory, MS 169-506, 4800 Oak Grove Dr., Pasadena, CA,91109, USA
[c] California Institute of Technology, Mail Code 320-47, 1200 E. California Blvd., Pasadena, CA, 91125, USA
[d] University of Chicago, Enrico Fermi Institute and Dept. of Astronomy and Astrophysics, Chicago, IL, 60637, USA
[e] University of Chicago, Enrico Fermi Institute, Chicago, IL, 60637

## ABSTRACT

We have developed a fore-optics module that converts the SHARC-II camera at the Caltech Submillimeter Observatory into a sensitive imaging polarimeter at wavelengths of 350 and 450 μm. We refer to this module as "SHARP". SHARP splits the incident radiation into two orthogonally polarized beams that are then re-imaged onto opposite ends of the 32 × 12 pixel detector array in SHARC-II. A rotating half-wave plate is used just upstream from the crossed grid. The effect of SHARP is to convert SHARC-II into a dual-beam 12 × 12 pixel polarimeter. A novel feature of SHARP's design is the use of a crossed grid in a submillimeter polarimeter. Here we describe the detailed optical design of SHARP and present results of tests carried out during our first few observing runs. At 350 μm, the beam size (9 arcseconds), throughput (75%), and instrumental polarization (< 1%) are all very close to our design goals.

## 1. INTRODUCTION

The earliest detections of far-infrared/submillimeter polarization in astronomical objects were obtained during the 1980s using single-pixel polarimeters operated in the stratosphere from balloons[1] and aircraft[2]. This work established a new technique for mapping interstellar magnetic fields. During the 1990s, astronomers developed more powerful polarimeters with many tens of pixels. Two examples built at the University of Chicago are Stokes[3] for the Kuiper Airborne Observatory (KAO), and Hertz[4,5] for the Caltech Submillimeter Observatory (CSO) on Mauna Kea. These instruments gathered polarization data for more than 50 star forming clouds at up to several hundred sky positions per cloud, finding that dust emission is measurably polarized at almost every point[5,6,7,8]. Another multi-pixel polarimeter developed during the 1990s was SCU-POL[9], the polarimeter for the Submillimeter Common-User Bolometer Array[10] (SCUBA) used with the James Clerk Maxwell telescope (JCMT) on Mauna Kea. SCU-POL was used to map magnetic fields in both high and low mass star-forming regions[11], and to obtain the first detection of submillimeter polarization in circumstellar disks around T-Tauri stars[12].

In the early years of the present decade, submillimeter polarimetry has been extended to both larger and smaller angular scales. For example, the Submillimeter Array[13] (SMA) on Mauna Kea studied the Sagittarius A* source at the Galactic center with sub-arcsecond resolution[14], while Northwestern University's Submillimeter Polarimeter for Antarctic Remote Observations[15] (SPARO) mapped the degree-scale magnetic fields in the Galactic center[16] and in Giant Molecular

---

[*] Presently at The Harvard Smithsonian Center for Astrophysics, 60 Garden Street, Cambridge, MA 02138, USA

Clouds in the Galactic disk[17]. Cosmologists are interested in millimeter-wave polarimetry as a probe of the early universe[18], and as a byproduct they have produced high quality maps of the global magnetic field of the Galaxy[19, 20].

The polarimeters discussed above were developed using varying approaches. In some cases, polarimetric capability was introduced by incorporating new optical elements into detector systems already in place. In other cases, new polarimeters were built complete with their own detectors. The former approach has obvious advantages in terms of cost, and often involves shorter lead times.

Another difference involves dual- versus single-beam polarimetry. A dual-beam polarimeter simultaneously detects two orthogonal components of polarization, while a single-beam system detects only one component at a time. The former approach avoids wasting photons. Furthermore, if the noise affecting orthogonal components is correlated[5], then the dual-beam capability can provide for very large improvements in sensitivity due to noise cancellation. Specifically, an important noise source affecting submillimeter observations is variability in atmospheric emission on short time scales, referred to as "sky noise". At the shorter submillimeter wavelengths (the 350 and 450 μm atmospheric windows) this effect is especially severe, and the sky noise affecting orthogonal polarization components is generally highly correlated. Thus, the dual-beam design is especially attractive for these shorter wavelengths.

SHARP is a fore-optics module that converts the CSO's SHARC-II[21] (Submillimeter High Angular Resolution Camera generation II) into a dual-beam polarimeter. As shown in Figure 1, the SHARP module is inserted into the optical train ahead of the SHARC-II cryostat. Like SHARC-II, SHARP, can be operated at either 350 μm or 450 μm. The incoming beam is split into orthogonal components of polarization that are directed to opposite ends of the $32 \times 12$ pixel bolometer array (Fig. 2), where the two components are recorded simultaneously. A rotating half-wave plate is located just upstream of the point where the beam is split. The data acquisition scheme involves performing standard photometric integrations at each of four half-wave plate rotation angles (0°, 22.5°, 45°, and 67.5°) successively. The photometric integrations have been carried out in chop/nod mode[7, 22, 23], but we plan to use scanning mode[21] in the near future. SHARP is the first dual-beam submillimeter polarimeter developed by adding polarimetric capability to an existing camera.

In this paper we describe the optical design of SHARP (section 2) and present results of tests carried out during the first 18 months of operation at CSO (section 3). In previous papers we described an early version of the optical design[24], and our first-light tests[25]. Initial scientific results will be described in future papers.

## 2. OPTICAL DESIGN

### 2.1 Overview

In addition to achieving dual-beam capability, our major goals for the design of SHARP were that (1) the SHARP installation should not require moving or altering any existing optical elements in the CSO/SHARC-II optical train; and (2) the spatial resolution and sensitivity of SHARC-II should be maintained. To achieve these goals, it was necessary to increase the optical path to make room for the new optical components, while at the same time maintaining the location of the final focal plane at the SHARC-II detector array. This required the use of re-imaging optics.

Figure 1 is a schematic drawing of the optical interface between the CSO telescope and SHARP, with the location of the SHARP module also shown. SHARC-II uses the Nasmyth focus of the CSO, but the camera does not sit directly at the Nasmyth focus. Instead, this focus is re-imaged along a folded optical path ~2.5 m in length that includes a flat mirror (M4, Fig. 1) and an ellipsoidal mirror (M5, Fig. 1). The SHARC-II bolometer array is located at the re-imaged focal plane, which is faster than the Nasmyth focus. (The final focal ratio is about 4, which is about three times smaller than the Nasmyth focal ratio.) An image of the primary (a "pupil") is located several inches in front of this focus. SHARC-II has a cold stop at this pupil. SHARP is installed into the limited physical space between the hollow elevation bearing and the M4 mirror (see Figs. 1, 4 and 5).

Another design goal was that high polarization efficiency should be achieved for both SHARC-II passbands (350 and 450 microns) with the capability to easily switch between those passbands.

## 2.2 Focal plane re-imaging

The optical diagram in figure 3 shows an "unfolded" version of the SHARP module, i.e., a version in which all flat mirrors and polarizing grids have been omitted. Although this diagram fails to capture the important polarization-splitting function of our module, it serves to illustrate the use of "crossed paraboloids" for re-imaging with minimal aberration[26]. A pair of identical paraboloidal mirrors, located immediately downstream from the Nasmyth focus, serves to re-image the Nasmyth focus from position f to a new position f′. The optical path between f and f′ is then used for the polarization-splitting optics. By making a judicious choice for the location of the final focus f′, we eliminate the need to move any component of the SHARC-II system when installing SHARP (see section 2.3).

Two important elements of the re-imaging design are that the angles of incidence at the two paraboloids are identical, and that the pupil is located approximately half-way between the two paraboloids[26]. Another important consideration is primary illumination. We will discuss this issue in section 2.4, after first describing the optical path in section 2.3.

## 2.3 Optical path

The arrangement of the polarimeter components is illustrated in Fig. 4. After the first paraboloidal mirror (P1) and the flat mirror (F1), incident radiation reaches the half-wave plate (HWP). This is mounted in a "half-wave plate module" (Fig. 6) which contains two interchangeable crystal quartz half-wave plates, one for each of the SHARC-II passbands at 350 and 450 μm, respectively. The HWP is rotated at regular intervals during observations (see section 1). The HWP is located near a pupil, (about half-way between P1 and P2 in Fig. 3) in order to minimize the size of the plate (diameter ~ 10 cm). This location also ensures that different parts of the field-of-view use nearly the same portion of the plate.

Also near the pupil, for the same reasons, is the crossed grid[27] (XG), which divides the beam into two orthogonal polarization components. The XG consists of two free-standing wire polarizing grids that interpenetrate one another, intersecting at a right angle. One grid has wires running parallel to the line of intersection and the other grid has wires running in the perpendicular direction. The effect of the XG is to direct the two orthogonal components toward opposite directions, both perpendicular to the incident beam (Fig. 4). The crossed-grid represents a compact and symmetrical solution to the problem of polarization-splitting, which was advantageous for SHARP due to the limited physical space avaialable. The crossed-grid was constructed by QMCI (Cardiff, Wales).

After exiting the XG, each polarization component is reflected by a pair of flat mirrors: F2h-F3h for the horizontally polarized (h) component and F2v-F3v for the vertically polarized (v) component. Because the h and v components travel along separate optical paths after the XG, we need two copies of paraboloidal mirror P2 (P2h and P2v). The converging beams from P2h and P2v next reach the beam combiner (BC) after 90° reflections by grids Gh and Gv, respectively. (The reason for using grids instead of mirrors here will be explained in section 2.5.) The BC is composed of two flat mirrors which deflect the corresponding beams by approximately 90° in the horizontal direction. (This reflection is discussed further in section 2.4 below.) After two more reflections by flat mirrors (F4 and F5), the two beams are directed toward SHARC II. The Nasmyth focus is re-imaged at the beam combiner, where the optical distance to the detector array is the same as that from the original Nasmyth focus in the original SHARC-II optical path.

After the BC, the h and v beams, originally from the same part of the sky, are displaced horizontally so that they are imaged on different halves of the detector array (Fig. 2). For the original chief ray of SHARC-II, the separation of the h and v beams is 20 pixels on the detector array. We do not use the central $8 \times 12$ pixels which might be reached by both polarization components in case of misalignment, diffraction, etc. Thus, the h and v subarrays each include $12 \times 12$ pixels. The resulting SHARP field of view is ~ 55″ × 55″.

The optical elements shown in Figure 4 are enclosed within four separate aluminum boxes, as can be seen in Figure 5. This modular design facilitates installation and also makes it easy to change between camera mode and polarimeter mode, by removing "Box 4", which contains P1 and F5.

**2.4 Correct primary illumination**

The design discussed above has the effect of re-imaging the Nasmyth focal plane precisely onto the SHARC-II detectors. However, this is not a sufficient condition for the system to be optimized. One also needs to ensure that the time reversed rays from all SHARC-II pixels properly "illuminate" the primary mirror. For the SHARC-II design, a cold pupil inside the SHARC-II camera (located at an image of the primary) serves this purpose. Acting together, M5, M4, M3 and M2 (the secondary) correctly image this cold pupil onto the primary mirror, such that the edges of the stop correspond to the edges of the primary, for all pixels.

By tuning two parameters of the SHARP design, we were able to achieve correct primary illumination for SHARP. These parameters are the distance between P1 and P2 (see Fig. 3) and the angle by which the beams are deflected at the beam combiner (BC; see Fig. 4). The angle at the BC is precisely set to 89.8 degrees, rather than 90, by fabricating it as a roof–shaped optical surface from a single block of aluminum alloy.

**2.5 Unused polarization components**

The light from the sky that enters the XG after passing through the HWP is not the only radiation entering the XG. For example, noise (thermal emission from Box 3) enters the XG from the opposite direction and the XG divides this noise into two polarized components that travel, together with the astronomical signals, along the two different paths leading from the XG down to the BC. This noise must be prevented from reaching the detectors. Note that this noise will be polarized orthogonally to the signals from the sky. The reason for using grids at Gh and Gv (Fig. 4), rather than mirrors, is to reflect only the astronomical signals while filtering out the noise that gets into the system from the "wrong" side of the XG.

Gh and Gv alone will not solve the problem, because the detectors can still "look" through these grids and see the emission from anything located behind the grids. Gh and Gv are coupled with a pair of on-axis paraboloidal mirrors (P3h and P3v, Fig. 4), which we refer to as "cold load mirrors" (Fig. 5). The focal length of the cold load mirrors was chosen such that, through them, the detectors will look back into the cryostat, where the thermal emission is much lower. Thus, these "unused polarization components" should not add appreciable excess noise.

## 3. PERFORMANCE

SHARP was first operated at the CSO in August 2005. Subsequent SHARP observations were made in January, July, and December of 2006. Here we describe the performance of SHARP as determined from data collected during the three 2006 runs. Science results obtained will be described in separate papers. All test and science data acquired in 2006 were collected using the chop-nod mode (see section 1) with the exception of the beam-size measurements described in section 3.1, that were made using both chop-nod and scanning modes.

We also include a very brief discussion of the sensitivity of SHARP (section 3.5). The sensitivity depends critically on the data analysis techniques, which perform the crucial task of removing the effects of sky noise (see section 1). SHARP cannot cancel the effects of sky rotation as has been possible for previous dual-beam submillimeter polarimeters, such as Hertz and STOKES, which employed instrument rotators. Thus, the analysis techniques that were developed for these instruments[7, 22, 23] are not directly applicable. The new techniques that we developed for SHARP, and the extent to which they have been successful in removing sky noise, will be described in a future paper.

**3.1 Beam size**

For operation at 350 μm, SHARC-II achieves a beam full-width at half-maximum (FWHM) of 8-9 arcsec. During July 2006, we acquired images of Uranus using SHARP in scanning mode, which involves slowly scan the telescope without chopping[21] (see section 1). The result, shown in Figure 7, gives a beam FWHM of ~ 9 arcsec. In December 2006, we used the same method to measure a beam FWHM of ~ 10 arcsec at 450 μm.

During July 2006, we also obtained 350 μm images of Uranus in chop-nod mode, and measured a beam FWHM of ~ 10 arcsec. This increase is probably due to jitter in the position of the secondary mirror during data acquisition.

Figure 7 shows that the v-sub-array obtains a slightly elongated beam, with one axis ~15% longer than the other. This is due to the construction of the crossed-grid (XG). Recall that the XG consists of two intersecting grids having vertical and horizontal wires, respectively. In fact, the "grid" having vertical wires is composed of two separate grids, one on each side of the horizontal grid. These have been carefully aligned by the manufacturer so that they produce approximately the same effect as a single uninterrupted grid. However, at the point where the two separate vertical grids meet, there is a translational misalignment of about 0.001 inch. The result is a slight elongation of the v-beam.

### 3.2 Throughput

For each sub-array of SHARP, ten reflections have been added to the original CSO/SHARC-II optical path. Thus one can expect that the fraction of the incident photons from a given sky position that are absorbed at the detector array will be smaller for SHARP observations than for SHARC-II observations. One can predict this loss based on the properties of SHARP's optical elements. Here we carry out such a prediction, for the 350 μm passband. All SHARP mirrors have been gold-coated except for P1, P2, and BC, which are machined from Aluminum alloy. Thus absorption losses should be below 0.5% per surface. The manufacturer of the paraboloidal mirrors P1 and P2 has estimated the rms surface errors to be of order 5 μm, which implies about 3% scattering loss for each paraboloid[28]. All flat mirrors are of optical quality so scattering losses should be negligible at submillimeter wavelengths. We expect an additional 5% loss due to imperfect performance of polarizing grids. The largest source of inefficiency is the half-wave plate (HWP), which is expected to absorb 5-10 % of the radiation[29] and to cause reflection losses of 1% per surface due to the non-ideal behavior of the anti-reflection coatings. We sized all of the optical elements to preserve five Airy rings. ZEMAX simulations show that this should reduce diffraction losses to below 1%. Thus, the resulting optical efficiency, or throughput, of SHARP relative to SHARC-II is expected to be about $0.995^{10} \times 0.97^2 \times 0.95 \times 0.925 \times 0.99^2 = 77\%$.

We measured the SHARP throughput at 350 μm by observing Mars with and without SHARP installed. Comparing the integrated flux (summed over all pixels) of these two modes yields an efficiency of 75%, in good agreement with the predicted SHARP throughput. Note that when SHARP is installed the radiative loading on the bolometers decreases, causing the bolometer responsivity to increase. We have corrected for this effect when calculating the above reported efficiency.

### 3.3 Instrumental polarization (IP)

During our January 2006 run, we observed Mars, Saturn, and Jupiter, which we used to estimate the IP under the assumption that all three planets are unpolarized. The IP is defined as the spurious polarization due to telescope and instrument, and in this paper the IP will be measured with respect to a reference frame that is fixed to SHARP. Specifically, the angle of polarization $\phi$ of the IP will be defined as zero for vertical polarization, and increase counter-clockwise as viewed by an observer standing at the position of SHARP and looking toward M3 (see Fig. 1).

We model the IP as the sum of two components. One is fixed with respect to SHARP itself (the "fixed component") and is presumed to be caused by the SHARP module. The other component is fixed with respect to M3 (the "M3 component") and is presumed to be due to polarization induced by the reflection at the M3 mirror (see Fig. 1). Due to the Nasmyth location of SHARC-II, the angle of the M3 component will depend on the telescope elevation.

M3 is a front-surface aluminized glass plate. At submillimeter wavelengths, reflection from aluminum mirrors is expected to be polarized at the level of several tenths of a percent, in a direction perpendicular to the plane defined by the incident and reflected rays[30]. Thus, the M3 component of the IP is expected to be vertical when the telescope is pointed at the horizon (Fig. 1).

Our IP model has four free parameters. These parameters are the magnitude $P_f$ and angle $\phi_f$ of the fixed component, and the magnitude $P_{M3}$ and angle offset $\delta\phi_{M3}$ of the M3 component. The angle offset $\delta\phi_{M3}$ is defined such that the angle of the M3 component is equal to $\varepsilon + \delta\phi_{M3}$, where $\varepsilon$ is the telescope elevation. From the above discussion we can see that the theoretical expectation is $\delta\phi_{M3} = 0°$. The normalized Stokes parameters of the total IP can then be written as:

$q = P_f \cos(2\phi_f) + P_{M3} \cos[2(\varepsilon + \delta\phi_{M3})]$, and

$u = P_f \sin(2\phi_f) + P_{M3} \sin[2(\varepsilon + \delta\phi_{M3})]$

From a least squares fit to the January 2006 planet observations, we obtain: $P_f = 0.46\%$, $\phi_f = 103°$, $P_{M3} = 0.42\%$, and $\delta\phi_{M3} = 1°$. Note that this value for $\delta\phi_{M3}$ agrees well with the theoretical expectation. From the theory of the classical skin effect, $P_{M3}$ should be several times smaller than what we have measured, but this theory has been shown to significantly underestimate the polarization magnitude for real mirrors at submillimeter wavelengths[30]. Regarding the fixed component, possible explanations include polarization by reflection at P1 and F1 as well as non-ideal behavior of the half-wave plate. The total IP is always below 1%, regardless of elevation.

During our July 2006 run, we observed Jupiter only. The results from a least-squares fit to these data are $P_f = 0.43\%$, $\phi_f = 177°$, $P_{M3} = 0.29\%$, and $\delta\phi_{M3} = 0°$. The only significant change is in the angle of polarization of the fixed component, which we attribute to the different zero angles of the extraordinary axis of the HWP that were used for the two runs. Preliminary analysis of SHARP data collected at 450 μm in December 2006 shows that the total IP is also below 1% for this passband.

### 3.4 Polarization efficiency

To measure the polarization efficiency we place a cold (~ 77 K) load between M3 and SHARP. The radiation from the cold load is then polarized nearly 100% by passing through two polarizing wire grids in series, with their wires parallel to one another. With this arrangement we measure the polarization efficiency to be 93% ± 1% at 350 μm and 97% ± 1% at 450 μm.

### 3.5 Sensitivity for chop-nod mode

Using the SHARP characteristics discussed above, we can predict the sensitivity of SHARP when used in chop-nod mode under the assumption that all sky noise is removed by the data analysis. In this case, the measurement errors will be set by quantum fluctuations in the atmospheric background, as is generally true for total intensity measurements made with SHARC-II[21]. Background-limited errors in total intensity for typical "good weather" conditions ($\tau_{230\,GHz} = 0.05$; target at 1.3 airmasses) are straightforward to estimate[21]. To first order, we can assume that for SHARP observations of a Stokes parameter (Q or U), the background-limited errors corresponding to this polarized flux measurement will be equal to the background-limited errors in total flux that one obtains for a SHARC-II measurement having the same integration time. To see this, note that polarized flux measurements are obtained from differences between h and v signals, while total flux measurements are obtained from the corresponding sums, and recall that the propagated errors are identical for these two operations.

Errors in measured polarization for a given source flux and observing time can then be estimated by dividing polarized flux errors by source flux. In carrying out this calculation, we have taken care to (a) degrade the signal to noise by a factor of (1/0.75) to account for SHARP losses (section 3.2), (b) allow time to observe both Q and U, (c) use the degradation factor appropriate for chop-nod mode[21], and (d) allow for a difference between observing time (which includes settle time for chopper, telescope, and half-wave plate, file overheads, and time for pointing and calibration) and integration time (which includes only time spent collecting data while pointed at source or reference positions). Assuming an observing efficiency (ratio of integration time to observing time) of 50%, this calculation gives a required 350 μm point source flux of 2.6 Jy for a ± 1% polarization measurement in 5 hours of observing time. (We have assumed here that the unused polarization components do not add noise; see section 2.5.)

We are still optimizing the SHARP observing efficiency, which was 25% in July 2006. We expect that we can reach 50%. A preliminary analysis of DR 21 observations carried out during marginal weather ($\tau_{230\,GHz} = 0.08$) in July 2006 shows that SHARP achieved background limited signal-to-noise.

## 4. SUMMARY

We developed an optical module, SHARP, that converts the CSO submillimeter camera SHARC-II into a dual-beam polarimeter. The design uses a HWP to modulate the polarization signal, a crossed-grid as a polarizing beam splitter, and crossed paraboloids to re-image the original SHARC-II focal plane. The compact modular design makes it simple to install and remove, which allows for rapidly switching between camera and polarimeter modes during a single observing session. Additionally, there is no degradation of the optical performance of the camera. SHARP operates with high polarization efficiency at both of SHARC-II's primary passbands of 350 and 450 μm and has low (< 1%) instrumental polarization. At 350 μm, the throughput is 75%.

Funding for SHARP has been provided by NSF awards to Northwestern University and to The University of Chicago, and by NSERC awards to The University of Western Ontario. We thank SHARP collaborators Roger Hildebrand, Megan Krejny, Martin Houde, Michael Attard, Hiroko Shinnaga, Jacqueline Davidson, Lerothodi Leeuw, David Chuss, Alex Lazarian, and Farhad Yusef-Zadeh for contributing to proposals, carrying out observations, and working on data analysis. We thank Richard Wylde and Paul Waltz for help with the design of SHARP, and Richard Chamberlain and the staff of the Caltech Submillimeter Observatory for assistance at the telescope. Finally, we are very grateful to Thang Bui and to the staff of the N.U. Instrument Shop for carrying out the mechanical design and fabrication of SHARP.

**REFERENCES**

1. Cudlip, W., Furniss, I., King, K.J., Jennings, R.E. 1982, "Far infrared polarimetry of W51A and M42", MNRAS, 200, 1169
2. Hildebrand, R.H., Dragovan, M., & Novak, G. 1984, "Detection of submillimeter polarization in the Orion Nebula", ApJ, 284, L51
3. Platt, S.R., Hildebrand, R.H., Pernic, R.J., Davidson, J.A., & Novak, G., 1991, "100-micron array polarimetry from the Kuiper Airborne Observatory - Instrumentation, techniques, and first results", PASP, 103, 1193
4. Schleuning, D.A., Dowell, C.D., Hildebrand, R.H., Platt, S.R., & Novak, G. 1997, "HERTZ, A Submillimeter Polarimeter", PASP, 109, 307
5. Dowell, C.D., Hildebrand, R.H., Schleuning, D.A., Vaillancourt, J.E., Dotson, J.L., Novak, G., Renbarger, T., & Houde, M. 1998, "Submillimeter Array Polarimetry with Hertz", ApJ, 504, 588
6. Hildebrand, R.H., Dotson, J.L., Dowell, C.D., Novak, G., Schleuning, D.A., Vaillancourt, J.E. 1998, "Hertz: an imaging polarimeter", in "Advanced Technology MMW, Radio, and Terahertz Telescopes", Ed. T.G. Phillips, SPIE Proc. Ser., 3357, 289
7. Hildebrand, R.H., Davidson, J.A., Dotson, J.L., Dowell, C.D., Novak, G., &Vaillancourt, J.E. 2000, "A Primer on Far-Infrared Polarimetry", PASP, 112, 1215. (Erratum: 112, 1621)
8. Dotson, J.L., Davidson, J.A., Dowell, C.D., Schleuning, D.A., & Hildebrand, R.H. 2000, "Far-Infrared Polarimetry of Galactic Clouds from the Kuiper Airborne Observatory", ApJS, 128, 335
9. Greaves, J. S. et al., 2003, "A Submillimeter Imaging Polarimeter at the James Clerk Maxwell Telescope", MNRAS 340, 353.
10. Holland, W. S.et al., 1999, "SCUBA: a common-user submillimetre camera operating on the James Clerk Maxwell Telescope", MNRAS, 303, 659.
11. B. C. Matthews, "Polarimetry and Star Formation in the Submillimeter", 2005, in Astronomical Polarimetry - Current Status and Future Directions, ASP Conf. Ser. 343, Eds. A. Adamson, C. Aspin, C. J. Davis, and T. Fujiyoshi (San Francisco: ASP) p. 99
12. Tamura, M. et al. "First Detection of Submillimeter Polarization from T Tauri Stars", 1999,ApJ, 525, 832
13. Moran, J. M., "Submillimeter array", 1998, SPIE, 3357, 208.
14. Marrone, D. P, Moran, J. M, Zhao, J.-H, Rao, R., 2007, "An Unambiguous Detection of Faraday Rotation in Sagittarius A*", 654, 57
15. Renbarger, T. et al., 2004 "Early Results from SPARO: Instrument Characterization and Polarimetry of NGC 6334", PASP 116, 415-424.

16. Novak, G. et al., 2003, "First Results from the Submillimeter Polarimeter for Antarctic Remote Observations: Evidence of Large-scale Toroidal Magnetic Fields in the Galactic Center", Ap J, 583, 83
17. Li, H., Calisse, P. G, Chuss, D. T, Griffin, G. S, Krejny, M., Loewenstein, R. F, Newcomb, M. G., and Novak, G., 2006, "Results of SPARO 2003: Mapping Magnetic Fields in Giant Molecular Clouds", ApJ, 648, 340
18. Kovac, J. M.; Leitch, E. M.; Pryke, C.; Carlstrom, J. E.; Halverson, N. W.; Holzapfel, W. L., 2002, "Detection of polarization in the cosmic microwave background using DASI", Natur., 420, 772
19. Benoit, A. et al. 2004, "First detection of polarization of the submillimetre diffuse galactic dust emission by Archeops", A&A, 424, 571.
20. Page, L. et al., "Three Year Wilkinson Microwave Anisotropy Probe (WMAP) Observations: Polarization Analysis", astro-ph/0603450
21. Dowell, C. D, et al. "SHARC II: a Caltech submillimeter observatory facility camera with 384 pixels", 2005, in Proc. SPIE 4855: Millimeter and Submillimeter Detectors for Astronomy, eds. T. G. Phillips & J. Zmuidzinas, 73.
22. Li, H., Calisse, P. G, Chuss, D. T, Griffin, G. S, Krejny, M., Loewenstein, R. F., Newcomb, M. G., and Novak, G., "Results of SPARO 2003: Data Analysis and Polarization Maps", 2005, in Astronomical Polarimetry - Current Status and Future Directions, ASP Conf. Ser. 343, Eds. A. Adamson, C. Aspin, C. J. Davis, and T. Fujiyoshi (San Francisco: ASP) p. 43.
23. Kirby, L.; Davidson, J. A.; Dotson, J. L.; Dowell, C.D.; Hildebrand, R. H., 2005, "Improved Data Reduction for Far-Infrared/Submillimeter Polarimetry", PASP, 117, 991.
24. Novak, G. et al. 2004, "A polarimetry module for CSO/SHARC-II", SPIE 5498, 278.
25. Li, H.; Attard, M.; Dowell, C. D.; Hildebrand, R. H.; Houde, M.; Kirby, L.; Novak, G.; and Vaillancourt, J., 2006 "SHARP: the SHARC-II polarimeter for CSO", SPIE, 6275, 62751H
26. Serabyn, E., 1995, "Wide-Field Imaging Optics for Submm Arrays", in ASP Conf. Ser. 75: Multi-Feed Systems for Radio Telescopes, eds. D. T. Emerson & J. M. Payne, 74
27. Payne, J. M., Lamb, J. W., Cochran J. G., Bailey, N. J., 1994, Proc. IEEE 82, 811-823, Plambeck, R, Thatte, N., Sykes, P. (1992): Proc 7th Int. Cryocooler Conf., Santa Fe NM, Nov. 1992
28. Ruze, E.J., 1953, "The effect of aperture errors on the antenna radiation pattern", Nuovo Cimento Suppl. 9, 364–380.
29. Murray, A. G., Flett, A. M., Murray, G., & Ade, P. A. R. 1992, "High efficiency half-wave plates for submillimetre polarimetry" Infrared Physics, 33, 113.
30. Renbarger, T., Dotson, L., J., Novak, G., 1998, "Measurement of submillimeter polarization induced by oblique reflection from aluminum alloy", Applied Optics, 37, 28.

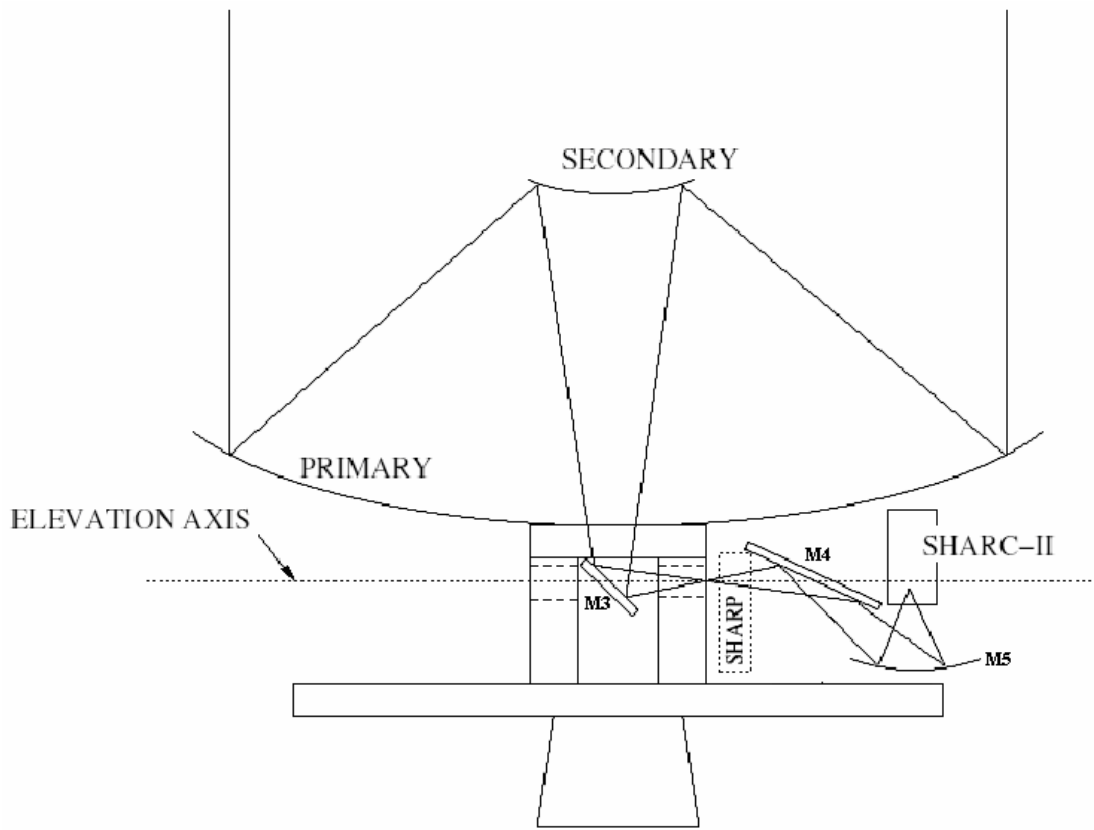

Fig. 1  Schematic drawing (not to scale) of SHARC-II on the Nasmyth platform, with the location of SHARP also shown. A flat mirror (M3) below the secondary deflects the incident beam into the hollow elevation bearing, producing an image of the sky within the bearing, at the Nasmyth focus. This focus is then re-imaged onto the SHARC-II detectors by mirrors M4 and M5. The removable polarimetry module SHARP is located between M3 and M4.

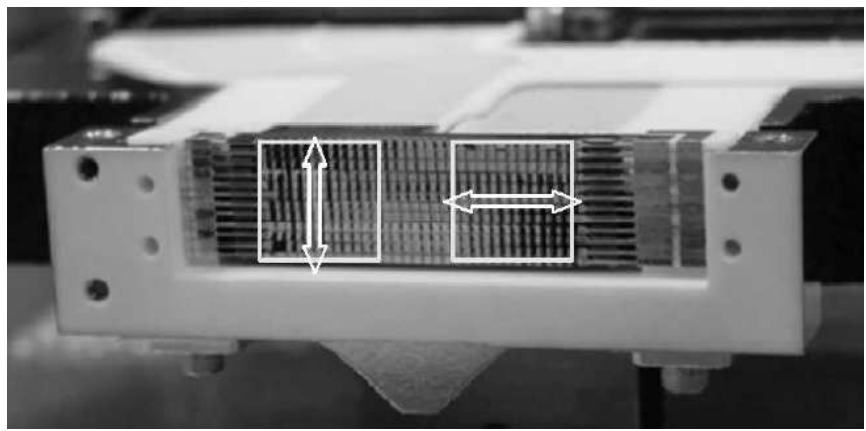

Fig. 2  Photograph of the SHARC-II detector array, with markings that illustrate the effect of the SHARP polarization-splitting optics. When SHARP is installed the 32×12 detector array is effectively converted into two 12×12 sub-arrays that view the same ~ 1′ × 1′ sky field in orthogonal polarizations.

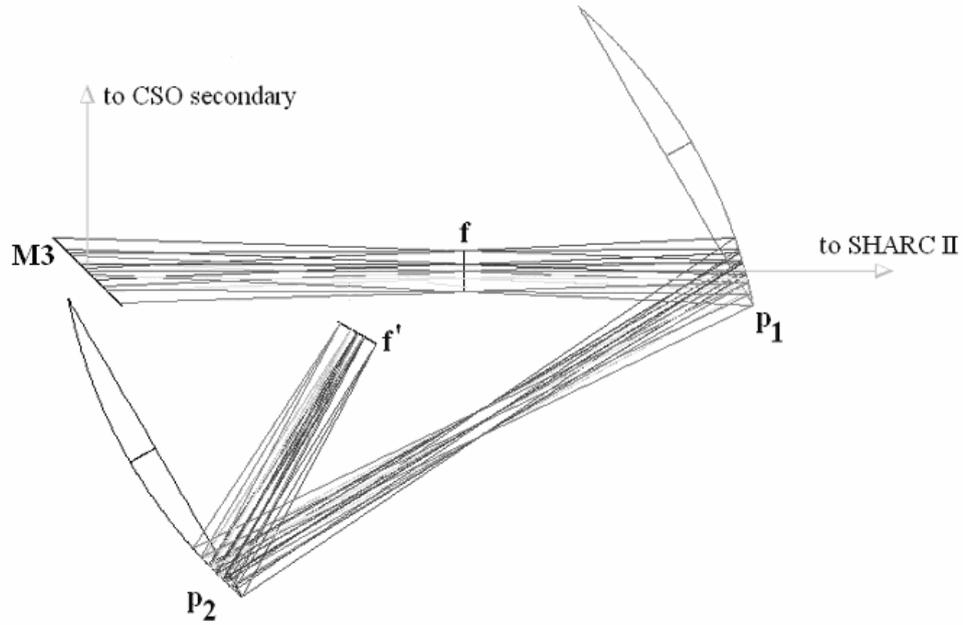

Fig. 3 "Unfolded" version of the optical design of SHARP, in which all reflections by flat mirrors and polarizing grids are omitted. Dual paraboloidal mirrors (P1 and P2) reimage the Nasmyth focus (f). The focus of the first mirror is placed on f, and f is re-imaged at f', the focus of the second paraboloid, with minimal aberration[26]. All the polarimetry components are installed after P1. The actual optical design of SHARP is illustrated in figure 4.

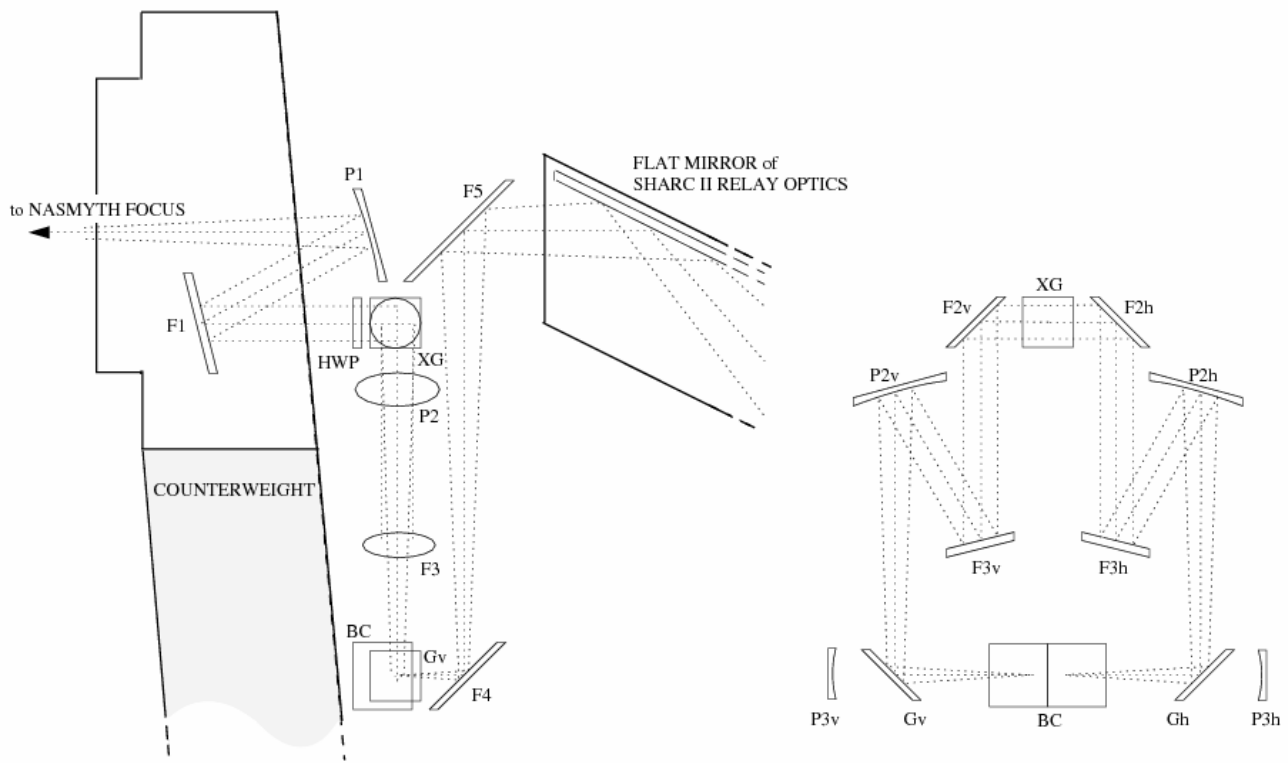

Fig. 4 Two views of SHARP. Left: The expanding beam from the Nasmyth focus is reflected by paraboloid P1 and by a flat mirror at F1, passes through the half-wave plate HWP, and reaches the crossed grid XG. From the XG, the horizontal polarization component propagates into the plane of the paper while the vertical component is directed towards the viewer. Right: View towards the Nasmyth focus. Vertical and horizontal components leaving the crossed grid undergo further reflections by mirrors and grids (F2v-F3v-P2v-Gv and F2h-F3h-P2h-Gh, respectively), ultimately bringing the components back together at the beam combiner BC which directs the recombined image toward the viewer. BC consists of two mirrors joined to form a roof–shaped optical surface. After reflection by BC, the two orthogonal polarizations are displaced laterally. The left view shows this reconstituted image being directed into the relay optics by flats F4 and F5. P1 and P2h (or P2v) form a pair of crossed paraboloids[26]. SHARC-II is easily converted back to photometric mode by removing P1 and F5 ("Box 4"; see Fig. 5).

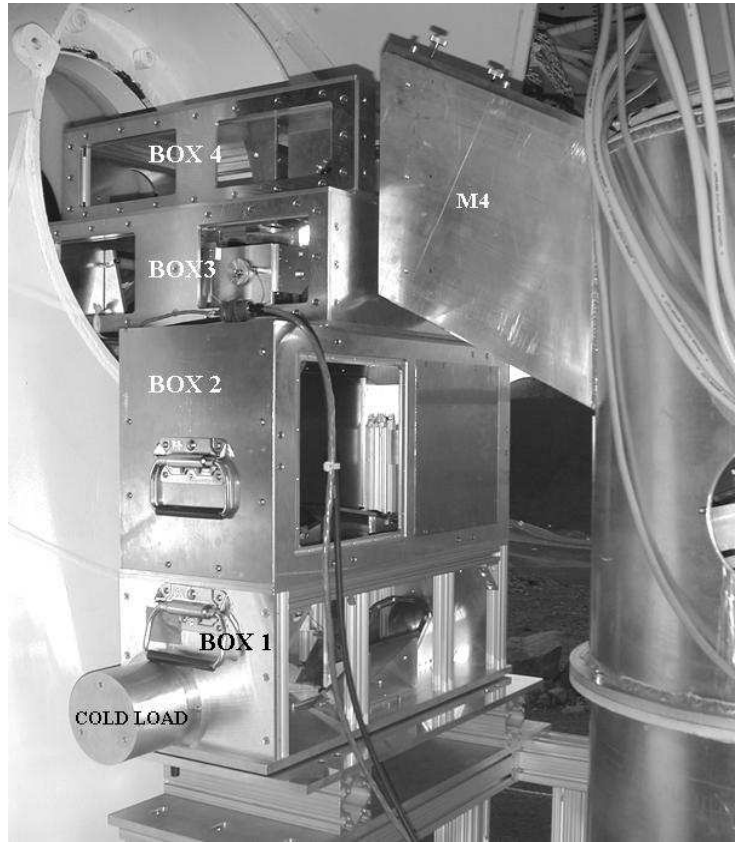

Fig. 5 The modular design of SHARP. The components in each box are: Box 1– P3h/v ("cold load" mirrors), Gh/v, F4, and BC; Box 2– P2h/v and F3h/v; Box 3– F1, HWP, XG, and F2h/v; Box 4 – P1 and F5.

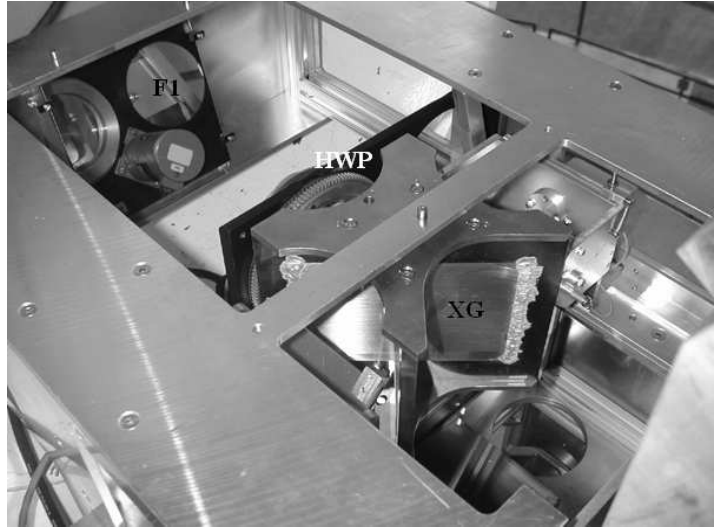

Fig. 6  F1, HWP, and XG in Box 3.  Box 4 was removed when this photo was taken.  F1 is the large square mirror, where the reflection of the half-wave plate module (HWP) can be seen. This reflection shows the stepping motor and the mounting positions for the two half-wave plates (only the 350 μm plate was installed in this picture).  The incident radiation passes through the HWP toward the XG where it is divided into two orthogonal polarization components.

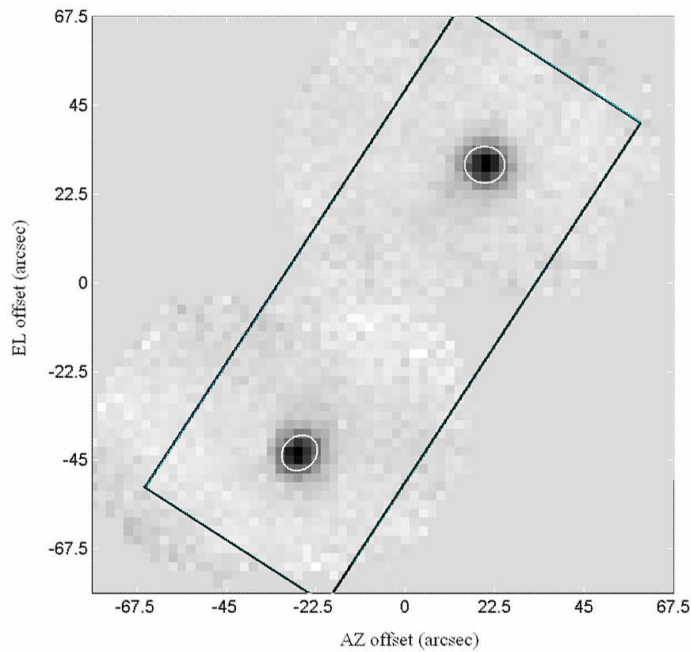

Fig. 7  A SHARP observation of Uranus at 350 μm.  Since Uranus is a point source at SHARP's resolution, this represents a measurement of the beam shape and size.  The observation was made in "scan mode" (see section 3.1) during July 2006.  The image at lower left is from the v-sub-array with the h-sub-array at upper right.  The black rectangle indicates the instantaneous field of view of SHARC-II and the white ovals indicate the beam FWHM values derived from two-dimensional Gaussian fits. The mean of the four FWHM values measured (two axes for each image) is 9.2 arcseconds.